\documentclass[smallextended]{svjour3}       
\smartqed  
\usepackage{graphicx}
\usepackage{dcolumn}
\usepackage{bm}
\usepackage{bbold}
\usepackage{braket}
\usepackage{url}
\usepackage{verbatim}
\usepackage{lineno}
\usepackage{amsmath}
\usepackage{amssymb}
\usepackage[utf8]{inputenc}
\usepackage{color}
\usepackage{lineno}
\usepackage{hyperref}
\usepackage{cite}
\usepackage[numbers,sort]{natbib}
\usepackage{rotating}

\newcommand{\asymland}{\rotatebox[origin=c]{90}{$\geqslant$}}

\begin{document}



\title{Relational Quantum Mechanics and Probability} 



\author{M. Trassinelli}
\institute{Institut des NanoSciences de Paris, CNRS, Sorbonne Université, 75005 Paris, France\\
\email{martino.trassinelli@insp.jussieu.fr}}


\maketitle

\begin{abstract}
We present a derivation of the third postulate of Relational Quantum Mechanics (RQM) from the properties of conditional probabilities.
The first two RQM postulates are based on the information that can be extracted from interaction of different systems, and the third postulate defines the properties of the probability function. 
Here we demonstrate that from a rigorous definition of the conditional probability for the possible outcomes of different measurements, the third postulate is unnecessary and the Born's rule naturally emerges from the first two postulates by applying the Gleason's theorem. 
We demonstrate in addition that the probability function is uniquely defined for classical and quantum phenomena. 
The presence or not of interference terms is demonstrated to be related to the precise formulation of the conditional probability where distributive property on its arguments cannot be taken for granted. 
In the particular case of Young's slits experiment, the two possible argument formulations correspond to the possibility or not to determine the particle passage through a particular path. 


\end{abstract}

\maketitle 

\section{Introduction}

After more than 100 years from the first attempt to formulate Quantum Mechanics (QM) by postulates \cite{Bohr1913}, its interpretation and foundation are still in discussion in the scientific community.
The origin of this open issue partially lies in the fact that our process of understanding is strongly based on the ability to make analogies with something familiar to us.
In the case of quantum phenomena, this approach fails because in our daily environment these kinds of events are not evident.
In the Young's slit experiment for example, where a massive particle or photons pass through the slits and form an interference pattern on a screen, we compare this phenomenon to common systems such as a ping-pong ball or sea waves.
The non-compatibility and contradictions between these two analogies create the impression to deal with paradoxes producing a sense of discomfort.
This difficulty to deal with quantum phenomena is well manifested by the famous phrase of R. Feynman who, in his sixth lecture  \textit{The character of physical law} at Cornell University in 1964 \cite{Feynman1967}, provocatively stated ``I think I can safely say that nobody understands quantum mechanics''. Without going into the discussion of the meaning of \textit{understanding}, different interpretations of quantum phenomena and different choices of QM postulates can be discussed. 
Several approaches have been formulated since the early days of Quantum Mechanics and have been amply discussed in the literature (see e.g. Ref.~\cite{Auletta} and references therein). 
New approaches and interpretations continue to be proposed 
\cite{Hardy2001,Clifton2003,Grinbaum2007,Grinbaum2007a,Dakic2009,Hohenberg2010,Chiribella2011,Fuchs2013,Masanes2013,Barnum2014,Dickson2015,Auffeves2016,Friedberg2018} demonstrating a strong interest in this subject still nowadays. 

To choose among the different QM interpretations and postulates formulations, a pragmatic approach is to consider two important criteria (i) the simplicity of the theory, following the Ockham's razor philosophical principle\footnote{From the English Franciscan friar William of Ockham statement \textit{``Non sunt multiplicanda entia sine necessitate''} , ``Entities must not be multiplied beyond necessity'' from William of Ockham (1287-1347), which can be interpreted in a more modern form as ``Among competing hypotheses, the one with the fewest assumptions should be selected''.}, and (ii) an interpretation that uses as ingredients for the postulates formulation, the concepts that are most familiar to our common experience.
To satisfy this second criterion, a way chosen by many authors in recent years is to use as basis of interpretations or foundation of QM the information that can be exchanged between the system and another system or an observer \cite{Rovelli1996,Brukner1999,Brukner2002,Fuchs2002,Clifton2003,Barrett2007,Brukner2009,Dakic2009,Chiribella2011,Masanes2013,Hohn2017,Hohn2017a,Hohn2017b,Yang2017,Yang2018,Yang2018a}.
The first postulation of QM in terms of information is due to Rovelli in 1996 with the formulation of Relational Quantum Mechanics (RQM)  \cite{Rovelli1996,Smerlak2007,Rovelli2018}.
In RQM framework, no system is privileged (there is no observable and observer, both are considered simply different systems) and everything depends on the reciprocal relationship through the mutual interaction, hence the name ``Relational'' in RQM.
The aspect of the quantization of nature is dictated in the first postulate by imposing a maximum limit on the information that can be extracted from a system. 
The probabilistic aspect of the theory is introduced by the second postulate that states that, even when all possible information is available, new one can be extracted from a system.
The third and last postulate, less intuitive, defines the way to compute probabilities, imposing, accordingly to experimental observations (Young's slits for ex.), the calculation via squares of amplitudes and sums of amplitudes (Born's rule).
In the context of RQM, new developments and discussions appeared in the last years \cite{Poulin2006,Grinbaum2007,Grinbaum2007a,Brown2009,vanFraassen2009}. In particular, EPR-type experiment have been reinterpreted in the context of RQM \cite{Smerlak2007}, it has been demonstrated that properties of orthmodularity are a direct consequence of the first two postulates of RQM \cite{Grinbaum2005} and, very recently, an alternative reconstruction of the Hilbert space is derived from the first two RQM postulates and four additional postulates \cite{Hohn2017,Hohn2017a,Hohn2017b}. 

In this article we present a derivation of the third postulate of RQM from the first two postulates and the basic properties of conditional probability function.
In the original article of RQM \cite{Rovelli1996}, Rovelli writes 
\begin{quotation}
\textit{``One could conjecture that equations (13)--(17) } (the equations relatives to probability properties, N.A.) \textit{could be derived solely by the properties of conditional probabilities \ldots Here, I content myself with the much more modest step of introducing a third postulate.''} 
\end{quotation}
The main result presented here is exactly to demonstrate Rovelli’s conjecture from first principles, thus eliminating the need of introducing a third postulate. 
We will demonstrate in particular that, when the most general probability function is rigorously defined, Born's rule emerges from the first two RQM postulates by the use Gleason's theorem.
The approach presented here is alternative to the recent work from Höhn \cite{Hohn2017,Hohn2017a,Hohn2017b} of QM foundation from the first two RQM postulated and additional ones.
Differences between the two methods and their corresponding assumptions will be discussed.

To present how to apply the developed formalism in a concrete case, we will consider the example of Young's slits experiment and we will calculate probabilities in cases of distinguishability or indistinguishability of the particle path through the slits pair.
We will in particular demonstrate that the probability function is uniquely defined for both cases.
The presence or not of interference terms is related to the precise formulation of the conditional probability where distributive property on its arguments cannot be taken for granted.

The article is organized as follows: in Sec.~\ref{sec:RQM}, as introduction, we present the first two RQM postulates and their relation to the measurement processes and Hilbert spaces. 
In Sec.~\ref{sec:prob} we discuss how we can built a probability function and we will study its properties.
Section~\ref{sec:slits} analyses the Young's slits experiment in the framework of the formalism developed.
We will end the article with the sections of conclusions (Sec.~\ref{sec:conclusions}) and an appendix for an introduction to orthomodular lattices and their connections with ``yes/no'' experiments.

\section{An introduction to Relational Quantum Mechanics: postulates, measurements algebra and Hilbert space} \label{sec:RQM}

This chapter is an introduction with some examples of RQM postulates. In particular, we recall here past major results on the theory of lattices associated to ``yes/no'' experiments and the reconstruction of the Hilbert space in the context of RQM. 

\subsection{Interacting systems and measurement processes}

The basic assumption of RQM is that the world can be decomposed into a collection of systems, each of which can be equivalently considered as an observing system or as an observed system.
The information is exchanged via physical interaction between systems.
The process of acquisition of information can be described as a ``question'' that a system (observing system) asks another system (observed system).
Since information is discrete, any process of acquisition of information can be decomposed into acquisitions of elementary bits of information by a series of ``yes/no'' experiment, i.e. a click or not of a specific measurement apparatus.\footnote{An answer to a question correspond then to one bit of acquired information.}

On another hand, a set of ``yes/no'' measurements on a same system preparation can be seen as a set of propositions linked to each other by an ordered relationship ``$\succeq$'' \cite{Birkhoff1936,VonNeumann,BeltramettiCassinelli,Hughes}.
If $i,j$ represent two possible measurement results, the relationship $i \succeq j$ indicates that each time we have a certain yes outcome for $j$ also we have a certain yes outcome for $i$.
The ensemble of propositions and  the order relation between them form a structure called \textit{partially ordered set (poset)} with some particular properties (orthocomplementation and other properties) that will be essential to define from it a Hilbert space.

It is not the intent of this article to make an introduction to poset and orthocomplemented lattices.
Here we will present only their basic properties in the relation with RQM postulates.
For the readers that are not familiar to these structures, their properties and their connection to ``yes/no'' measurements of classical and quantum phenomena, a short introduction is presented in the appendix at the end of the article.
An exhaustive presentation on this topic can be found in Refs.~\cite{BeltramettiCassinelli,Hughes}.

\subsection{First postulate of RQM and orthocomplemented lattices} \label{sec:1RQM}

The first postulate of RQM is \cite{Rovelli1996} 
 \begin{quotation} 
 \textit{Postulate 1 (Limited information): there is a maximum amount of relevant information that can be extracted from a system.} 
 \end{quotation}  
 If there is a maximal amount of information that can be extracted from the system, it exists an ensemble of $N'$ questions $Q^i$ that completely describes the system. 
 These questions can be represented by different experimental results: the detection of a particle with spin up or down or the measurement of a particle momentum within a value interval.  
 If the information is limited, this means that exists a list of limited ``yes/no'' questions $Q^{i}$ with $i=1,\ldots N \le N'$ independent from each other that describes completely the system. 
With the operations of conjunction ``$\land$'' (``AND''), disjunction ``$\lor$'' (``OR'') and negation ``$\lnot$'' (``NOT''), we can in addition build another list of mutually exclusive questions $Q_\alpha^{i}$ where $Q_\alpha^{i} \land Q_\alpha^{j} = \emptyset$ for $i \ne j$ \cite{Rovelli1996}, where the Greek letter indicate belonging to the same complete set of questions and ``$\emptyset$'' defines an empty set. 
For each pair of questions, the combinations $Q_\alpha^{i} \land Q_\alpha^{j}$ and $Q_\alpha^{i} \lor Q_\alpha^{j}$ are univocally defined.
Because we have a complete set of questions, we have that $\bigvee_i Q_\alpha^{i} = \mathbb{1}$, the ensemble of all possible positive answers, defines certainty (represented here by ``$\mathbb{1}$'').

For each question $Q_\alpha^{i}$ (e.g. ``is the spin up?'') its positive answer correspond to a statement $i$ (``the spin is up'').
A negative answer corresponds to its complementary statement $i^\bot$ (``the spin is not up''), which correspond to a positive answer to the question  $\lnot Q_\alpha^{i}$ opposite to $Q_\alpha^{i}$.\footnote{In other words we have $\lnot Q_\alpha^{i} = Q_\alpha^{i^\bot}$}
The different possible answers $\{i,i^\bot \}$ together with the operations ``$\land$'', ``$\lor$'' constitute a logical structure called \textit{lattice}, indicated by $\mathcal{L}$.
This lattice is orthocomplemented because for each element $i$ (a positive answer), it exist is complement $i^\bot$ (a negative answer to the same question), with the properties $i \lor i^\bot = \mathbb{1}$ and $i \land i^\bot = \emptyset$, where $\mathbb{1},  \emptyset$ are also part of the lattice $\mathcal{L}$ (a complete lattice).



Another important property of the lattice defined by a complete series of mutually exclusive questions is the distributivity, which corresponds to the property
\begin{equation}
i \land(j \lor k) = (i \land j) \lor (i \land  k),  \quad i \lor(j \land k) = (i \lor j) \land (i \lor k) \label{eq:distributivity}
\end{equation}
for any triplet $i,j,k \in \mathcal{L}$.
As we will see in the following sections where we will consider more than one set of complete questions, distributivity, or more precisely its absence, will play an important role to calculate interference phenomena in the probability calculation.
Note, a distributive orthocomplemented lattice is also called Boolean lattice.

\begin{figure}
\centering
\includegraphics[width=0.5\textwidth]{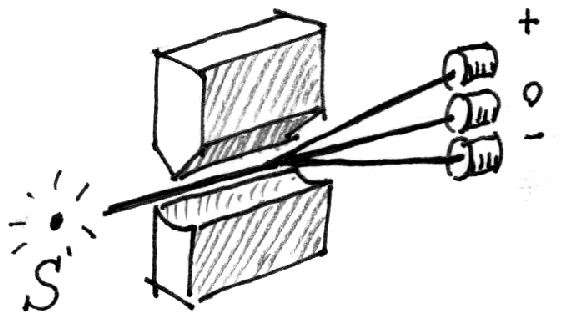} \hspace{1cm}
\includegraphics[height=0.25\textheight]{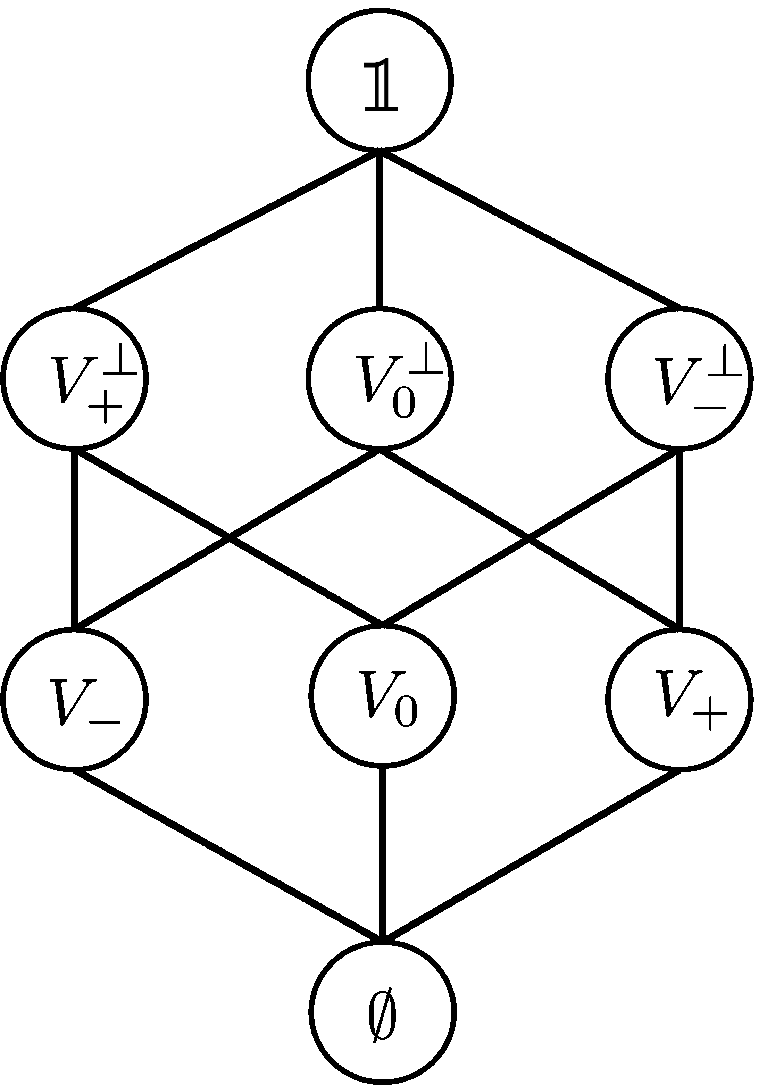}
\caption{Left: artistic scheme of a Stern-Gerlach experiment with vertically aligned magnets.
Right: graph of the orthocomplemented lattice corresponding to the detection of the three emerging beams.}
\label{fig:SG}
\end{figure}

A simple example of a complete set of questions can be built considering a spin-1 particle passing through a Stern-Gerlach apparatus with a defined axis orientation, vertical for example (see Fig.~\ref{fig:SG} left). 
Here the spin orientation of the particle is determined by the deflection of the particle itself passing through a non-homogeneous magnetic field.
The complete set of question is constituted by a series of three particle detectors that correspond to the questions $\{ Q_V \}= Q_V^+, Q_V^0$ and $Q_V^-$, one for each spin projection.
The ensemble of the corresponding answers $\{ V_+,V_0,V_-,V^\bot_+,V^\bot_0,V^\bot_-\}$ forms a orthocomplemented distributive (Boolean) lattice represented in Fig.~\ref{fig:SG} right (see Appendix for more details on lattices properties).

From the output of the three detectors we can extract only a limited amount of information.
In opposite, in a classical case as the measurement of egg holes by different calibers (see Appendix), from the use of additional holes with different diameters, additional information can be always be extracted.\footnote{If we do not consider the atomic structure of the eggs and calibers, the extractable information is in fact limitless.}

\subsection{Second postulate of RQM and orthomodular lattices} \label{sec:2RQM}

The second postulate of RQM is \cite{Rovelli1996}

\begin{quotation}
\textit{Postulate 2 (Unlimited questions):  it is always possible to acquire new information about a system.}
\end{quotation}

When a particle coming out from a Stern-Gerlach apparatus, vertically oriented for example, is injected into another Stern-Gerlach apparatus with a different orientation, e.g. horizontal, the new output can be used to extract new information with a loss of old information because its total quantity is bounded.
The new results $\{ H_+, H_0, H_-, H^\bot_+, H^\bot_0, H^\bot_-\}$ correspond to a new set of questions $\{ Q_H \}= Q_H^+, Q_H^0,Q_H^-$.
As for the vertical case, the new possible outcomes alone have a mutual relation similar to the scheme in Fig.~\ref{fig:SG} (right).
But when horizontal and vertical measurement apparatus are not considered subsequentially but together as alternative measurements, they form a lattice that is no more distributive, contrary to classical cases as a egg measured by two different sets of calibers (see Appendix).

The non-distributivity of the lattice is related to the fact that each set of questions is complete, but the sets are not compatible each other because each apparatus can extract an information $I_\text{max}$
and the maximal quantity of information that can be extracted is also $I=I_\text{max}$.
The absence of the distributive property has major consequences.
It implies in fact that the order on which the questions/measurements are made matters and is related to the commutation of operators in standard QM.

As pointed in the early years of QM \cite{Birkhoff1936}, proposition lattices defined by set of measurements of classical or quantum cases, have still in common an important property, \textit{orthomodularity}, which is a weaker property than distributivity\footnote{Orthomodularity derives from the modularity property, $i \lor(j \land k) = (i \lor j) \land k$ if $k \succeq i$ a special case of distributivity, and the property of orthocompleteness \cite{Birkhoff1936,BeltramettiCassinelli,Hughes}. All Boolean lattices, which are distributive, are automatically orthomodular.} consisting in 
\begin{equation}
i = j \lor (i \land  j^\bot) 
\end{equation}
for each pair of elements $i,j \in \mathcal{L}$ with  $i \succeq j$.
Recently, Grinbaum demonstrated \cite{Grinbaum2005} that orthomodularity property is in fact a direct consequence the presence of upper bound of information.
To note that for this demonstration the author introduces the property of \textit{relevance}\footnote{Question $j$ is \textit{irrelevant} with respect to the question $i$ if $j \land i^\bot \ne \emptyset$. \cite{Grinbaum2005}} between pairs of questions.
This property can be used to discriminate questions belonging to the same complete set or to different sets. 

A final comment to the second postulate is that it can be actually considered, as a lower bound on the information, in complement to the first postulate that determine the higher bound.
Absolute certainty corresponds to a zero information gain. Postulate 2 of RQM assure that we can extract always information larger than zero.

\subsection{Reconstruction of the Hilbert space}
It has been demonstrated that if we have an orthomodular lattice $\mathcal{L}$, which is in addition atomic, complete and has covering property, an Hilbert space $\mathcal{H}$ can be reconstructed \cite{BeltramettiCassinelli,Auletta,DallaChiara}.
As discussed above, orthomodularity can be derived directly from the first two RQM postulates \cite{Grinbaum2005}.
Similarly, completeness is directly derived by construction of  RQM because a system can be completely described by a finite number of answers of ``yes/no'' questions due to the information boundaries.
Moreover, each question constitutes actually an atom\footnote{Each positive (or negative) answer $i\ (i^\bot)$ to the question $Q_i$ is an elementary constituent and it respect the atomicity property: for any other element $j \in \mathcal{L}$, $i \succeq j$ implies that $j =\emptyset$ or $j = i$.} of the lattice $\mathcal{L}$. 
The covering property, valid for many lattices corresponding to measurements of quantum systems, has to be assumed valid the general case for the reconstruction of  $\mathcal{H}$ (on this assumption we will discuss more extensively in the next paragraphs).
Even if the Hilbert space $\mathcal{H}$ is reconstructed from the lattice formed by complete sets of questions, the choice of its numbered field (real, complex, quaternion, p-adic, etc. ) is not defined.
However, it can be demonstrated that the complex case satisfies the basic requirements for describing quantum phenomena without redundancy \cite{BeltramettiCassinelli}.

In the reconstruction of the Hilbert space from the orthomodular lattice, each complete set of questions $\{ Q_\alpha^{i}\}$ is associated to a basis $\{ \ket{i_\alpha}\}$ of $\mathcal{H}$.
Moreover, from any $\ket{i_\alpha}$ we can built a projector $P^i_\alpha = \ket{i_\alpha} \bra{i_\alpha}$.
Different sets of questions $\{ \ket{i_\alpha}\}, \{\ket{j_\beta}\}$ correspond to different complete bases that are related to each other via unitary matrices $U$ with $\ket{j_\beta} = U \ket{i_\alpha}$.
The unitarity of $U$ is guaranteed by the completeness and orthogonality of each basis element \cite{Cohen}.
In this context, a pure state, $\ket{i_\alpha}$, correspond to a certain positive answer to a question of a complete set.
A mixed state is the results of an uncertainty between answers of among questions.

\section{Probability function and properties} \label{sec:prob}

\subsection{Definition of probability function}

Taking into account two complete families of questions $\{ Q_\alpha^{i}\}, \{Q_\beta^{j} \}$ and the second RQM postulate, we can consider the probability for obtaining a positive answer to the question $Q_\beta^{j}$ knowing a positive answer to the question $Q^1_\alpha$. This is a perfect example of conditional probability that can be indicated by $p[Q_\beta^{j}|Q_\alpha^{i}]$.  In the next paragraphs we will build $p[Q_\beta^{j}|Q_\alpha^{i}]$ from minimal requirements for the most general definition of probability function. 

In its most general definition, probability is a measurement or map on a set of elements $\Omega$ to the real interval $[0,1]$ via a function $\wp: \Omega \to [0,1]$ \cite{SEPprob,Fine}. The first axiomatization of the probability function $\wp$ is due to Kolmogorov in 1933 \cite{Kolmogorov,Fine,Jaynes} where $\Omega$ is on a field $\mathcal{F}$ that is closed under complementation with respect to $\Omega$ and union\footnote{if $A$ and $B \in \mathcal{F}$,  also $A \cup B \in \mathcal{F}$ and the complement $\bar A = \Omega - A \in \mathcal{F}$ ($\mathcal{F}$ is a $\sigma$-field).} and where  \begin{subequations} \label{eq:axK} \begin{align}  &\wp [A] \ge 0 \text{ for all $A \in \mathcal{F}$}, \label{eq:axK1}  \\ &\wp[\Omega] = 1, \label{eq:axK2} \\ &\wp[A \cup B] = \wp[A] + \wp[B] \text{ for all $A$ and $B \in \mathcal{F}$ with $A \cap B = \emptyset$}.  \label{eq:axK3} \end{align} \end{subequations} This general form is normally implemented in different cases of classical or quantum physics \cite{Fine,Gudder,Varadarajan,Hughes,Auletta,DallaChiara} and is more or less explicitly used in all approaches of foundations of QM. 

In the context of logic of $True/False$ values of sentences of a set $\mathcal{S}$, an analog definition of $\wp$ is \cite{SEPprob,Cox,Fine} 
\begin{subequations} \label{eq:axL}
\begin{align} 
&\wp [A] \ge 0 \text{ for all $A \in \mathcal{S}$}, \label{eq:axL1}  \\
&\wp[\text{\textit{True}}] = 1, \label{eq:axL2} \\
&\wp[A \land B] = \wp[A] + \wp[B] \text{ for all $A$ and $B \in \mathcal{S}$ with $A \land B = \emptyset$}.  \label{eq:axL3}
\end{align}
\end{subequations}
This is a more adapted set of axioms of $\wp$ to be directly implemented for probability measurements in orthomodular lattices $\mathcal{L}$ \cite{BeltramettiCassinelli} and RQM where sentences of $\mathcal{S}$ are positive or negative answers of questions.

To correctly define the probability function $p[Q_\beta^{j}|Q_\alpha^{i}]$ for obtaining a positive answer to a question $Q_\beta^{j}$ knowing a positive answer to a question of another family $Q_\alpha^{i}$, we have to introduce conditionality properties.
To underline that we are dealing with measurements that corresponds to answers to questions, in the previous expression and in the following paragraphs we implicitly indicates by $Q^i$ its corresponding positive answer and $\lnot Q^i \equiv Q^{i^\bot}$ the corresponding negative answer.
Taking inspiration from Refs.~\cite{Ballentine1986,Fine,Jaynes,Ballentine2016}, from Eqs.~\eqref{eq:axL} we can derive the axioms

\begin{subequations} \label{eq:ax}
\begin{align} 
&p[Q_\beta^{j}|Q_\alpha^{i}] \ge 0, \label{eq:ax1}  \\
&p \Big[ \bigvee_j Q_\beta^{j} \Big| Q_\alpha^{i} \Big] = 1, \label{eq:ax2} \\
&p[Q_\beta^{j} \lor Q_\beta^{k}| Q_\alpha^{i}] = p[Q_\beta^{j} | Q_\alpha^{i}]  + p[Q_\beta^{k} | Q_\alpha^{i}],  \label{eq:ax3}
\end{align}
\end{subequations}
for the probability relative to the set of questions $Q_\beta^{j}$ knowing a positive question $Q_\alpha^{i}$.
The second axiom states that the probability of \textit{certainty} is 1. It corresponds to Eqs.~\eqref{eq:axK1} and  \eqref{eq:axL1}, and it is equivalent to $p[Q_\beta^{j}|Q_\beta^{j}]$ =1. 
Because of the completeness of each set of questions, we have in fact $ \bigvee_j Q_\beta^{j} \equiv \mathbb{1}$, i.e. always \textit{True}.
Due to the mutual exclusivity of the questions $Q_\beta^{j} \land Q_\beta^{k} = \emptyset$ for $j \ne k$, Eq.~\eqref{eq:ax3} is equivalent to Eq.~\eqref{eq:axL3}.

Similar definitions of conditional probability are implemented in the recent approach of QM foundation of Auffèves and Grangier \cite{Auffeves2016,Auffeves2017} and of Höhn \cite{Hohn2017,Hohn2017a,Hohn2017b} with however some important differences.
Höhn assumes additional symmetric properties of $p$ (see next paragraphs) to reconstruct QM framework.
This is not the case for Auffèves and Grangier where however the probability function is not rigorously defined, similarly to the case of RQM original formulation \cite{Rovelli1996}.

For a given answer of the question $Q_\alpha^{i}$ the probability function $p[ \cdot |Q_\alpha^{i}]$ defined by Eqs.~\eqref{eq:ax} is compatible with Kolmogorov's axiomatic definition (Eqs.~\eqref{eq:axK}).
Differently from Kolmogorov approach, we will see that the explicit dependency on the prior answers provides a clear use of $p$ in different contexts.

\subsection{The Gleason's theorem and the third principle of RQM} \label{sec:gleason}

In the previous sections we saw that from the first and the second postulates of RQM we can build a Hilbert space. This reconstruction enable to make a direct link to the different choices of probability axioms, Eqs.~\eqref{eq:axK} and \eqref{eq:ax} and then the use of Gleason's theorem \cite{Gleason1957}. 
Gleason's theorem (from an adaption of Refs.~\cite{Auletta,BeltramettiCassinelli}) states:
\begin{quotation}
\textit{
In a separable Hilbert space $\mathcal{H}$ of dimension $\ge 3$, whether real, complex or quaternion,
every probability measure} (that respects Eqs.~\eqref{eq:axK}, N.A.) \textit{on a closed subspace $\mathcal{H_S}$ can be written in the form
\begin{equation}
p[H_S] = tr[\rho P_\mathcal{S}],
\end{equation}
where $P_\mathcal{S}$ denotes the orthogonal projection on $\mathcal{H_S}$ and $\rho$ is a density operator.}
\end{quotation}

In our notation a pure state $\ket{ i_\alpha}$ corresponds to a certain positive answer to a question $Q_\alpha^{i}$ and it is associated to the density matrix $\rho_\alpha = \ket{ i_\alpha}\bra{ i_\alpha}$, which is in this case is equivalent to the projector $P^i_\alpha = \ket{ i_\alpha}\bra{ i_\alpha}$.

The probability for having a positive answer to the question $Q_\beta^{j}$ is
\begin{equation}
p[Q_\beta^{j}|Q_\alpha^{i}]  = tr[\rho_\alpha P_\beta^j] = | \braket{j_\beta | i_\alpha}|^2  = |U_{i j}|^2, \label{eq:gleasonQ}
\end{equation}
where $U_{ij}$ is the unitary matrix element corresponding to the transformation between the complete bases $\{ \ket{i_\alpha} \} $ and $\{ \ket{j_\beta} \}$ complete bases transformation and $P_\beta^j=\ket{ j_\beta}\bra{ j_\beta}$ is the projector corresponding to $Q_\beta^{j}$.
Thus, we see that Born’s rule can be deduced solely from the first two RQM postulates and from Gleason’s theorem, without the need of any further assumption: a third postulate devoted to probabilities does not appear necessary. 

From Eq.~\eqref{eq:gleasonQ} we see that $p[Q_\beta^{j}|Q_\alpha^{i}]$ is completely symmetric to the exchange of order of the questions $Q_\beta^{j}$ and $Q_\alpha^{i}$,
\begin{equation}
p[Q_\beta^{j}|Q_\alpha^{i}]  = p[Q_\alpha^{i} | Q_\beta^{j}], \label{eq:2Q}
\end{equation}
a property that is only assumed as valid in the original formulation of RQM \cite{Rovelli1996} but also in the recent QM foundation approach from Höhn \cite{Hohn2017,Hohn2017a,Hohn2017b} and in other recent works.

From Eqs.~\eqref{eq:ax}, we have in addition the properties
\begin{equation}
 p[\bigvee_j Q_\beta^{j}|Q_\alpha^{i}] = \sum_j p[Q_\beta^{j}|Q_\alpha^{i}] =  \sum_i  p[Q_\alpha^{i} | Q_\beta^{j}|] =  p[\bigvee_i Q_\alpha^{i} | Q_\beta^{j}|] = 1. \label{eq:sumrule}
\end{equation}
In other words $\sum_i |U_{ij}|^2 = \sum_j |U_{ij}|^2 =1$, which is one of the basic properties of unitary matrices \cite{Rovelli1996,Cohen}.

\subsection{Alternative foundations of QM via RQM postulates and discussion}

In the previous section we demonstrated that from minimal requirements for the conditional probability function and the first two postulates of RQM, Born's rule can be derived.
It has to be noted that this demonstration is based on the assumption that any lattice $\mathcal{L}$ associated to quantum measurements has covering properties \cite{BeltramettiCassinelli,Grinbaum2007}. 
This is the case for simple examples of simple quantum measurement graphs \cite{Birkhoff1936,Piron1972,Piron,BeltramettiCassinelli,Hughes} but it is not proved for the general case.
Here the cover property has to be assumed valid.\footnote{ A possible demonstration could maybe possible using information boundaries for lattice elements relative to different set of questions, similarly to the otrhomodularity derivation in Ref.~\cite{Grinbaum2005}.}
Moreover, the numerical field of the Hilbert space cannot be derived from RQM principles and real and complex numerical fields are found only from a non-exhaustive exclusion of other cases. 
In particular the quaternion case is eliminated only when composite systems are considered \cite{BeltramettiCassinelli}.

A recent alternative foundation of QM based on the first two principles of RQM, with addition of new postulates, has been proposed by Höhn \cite{Hohn2017,Hohn2017a,Hohn2017b}.
In this remarkable work, real and complex numerical fields are deduced by construction from a set of five postulates: the first two RQM ones plus other three.
The complex case becomes the only acceptable field once an additional postulate on the measurement of composite systems is added.
This is a very interesting result but it has to be noticed that, in addition to the important increase of number of postulates where time coordinate has a preferential role, it relies on the additional assumption of the symmetric properties of the probability functions $p[Q^i | Q^j]$ with respect couple of questions $Q^i,Q^j$. 
This is not the case in the demonstration proposed here were the symmetric properties of $p[Q^i | Q^j]$ are derived directly from Eq.~\eqref{eq:gleasonQ}.
It interesting to note that in Höhn's work, if the existence of a \textit{state of no information} is assumed, $p[Q_\beta^{j}|Q_\beta^{j}]$ is used to classify families of questions, \textit{independent, dependent, partially dependent}, similarly to the relation of \textit{relevance}  introduced by Grinbaum \cite{Grinbaum2005} between different lattice elements (see Sec.~\ref{sec:2RQM}).

\subsection{Additional properties of conditional probability function and Bayes' theorem} \label{sec:conditional}

Other non-trivial properties of the conditional probability function $p$ can be discussed considering three sets of complete questions.
If we have three subsequent specific questions $Q_\alpha^{i}, Q_\beta^{j}, Q_\gamma^{k}$ from three different sets of complete questions, we can consider the probability $p[Q_\gamma^{k} | Q_\beta^{j} \ \asymland \  Q_\alpha^{i}]$ for having a positive answer for $Q_\gamma^{k}$ knowing a positive answer for $Q_\beta^{j}$ \textit{after} a positive answer for $Q_\alpha^{i}$. 
To take into account the order of the questions we introduce the new non-symmetric relation operator ``$\asymland$'', similar to the ``AND'' logical operator ``$\land$'' but where $Q^i \ \asymland \ Q^j \ne Q^j \ \asymland \ Q^i$.
We underline the importance of the sequence of the questions because it is related to the non-commutativity of the operators $P_\alpha^{i}$ and $P_\beta^{j}$, i.e. $P_\alpha^{i} P_\beta^{j} \ne P_\beta^{j} P_\alpha^{i} $.
As we will see, when the sequence of the questions is not important, the operator ``$\asymland$'' reduces to the conjugation operator ``$\land$''.

From Gleason's theorem (Eq.~\eqref{eq:gleasonQ}) and the properties of the density operator, and in particular Lüders' rule for conditional probabilities, we have for pure states 
\begin{multline}
p[Q_\gamma^{k} |  Q_\beta^{j} \ \asymland \  Q_\alpha^{i}] 
 = \frac {tr[P_\beta^{j} \rho_\alpha P_\beta^{j} P_\gamma^{k}]}{tr[\rho_\alpha P_\beta^{j}]}
 = \frac {tr[P_\beta^{j} P^i_\alpha P_\beta^{j} P_\gamma^{k}]}{tr[P^i_\alpha P_\beta^{j}]}  =  \\ 
 = \frac {tr[P_\beta^{j} P^i_\alpha P_\beta^{j} P_\gamma^{k}]}{p[Q^i_\alpha |Q_\beta^{j}]}= \frac {| \bra{k_\gamma} P^j_\beta \ket{i_\alpha}|^2}{| \braket{j_\beta | i_\alpha}|^2}. \label{eq:3Q}
\end{multline} 
Because of the trace properties $tr[MNL]=tr[NLM]=tr[LMN]$ (but $\ne tr[MLN]$), the nominator  of Eq.~\eqref{eq:3Q} is completely symmetric with respect to the interchange of $P^i_\alpha$ and $P^k_\gamma$. 
The difference between $p[Q_\gamma^{k} |  Q_\beta^{j} \ \asymland \  Q_\alpha^{i}] $ and $p[Q_\gamma^{k} \ \asymland \  Q_\beta^{j} | Q_\alpha^{i}] $ is only the denominator, 
\begin{equation}
p[Q_\gamma^{k} \ \asymland \  Q_\beta^{j} | Q_\alpha^{i}] = tr[P_\beta^{j} P^i_\alpha P_\beta^{j} P_\gamma^{k}] = | \bra{k_\gamma} P^j_\beta \ket{i_\alpha}|^2,  \label{eq:3Qbis}
\end{equation}
which is required for the probability normalisation.
In other words, we have 
\begin{equation}
p[Q_\gamma^{k} |  Q_\beta^{j} \ \asymland \  Q_\alpha^{i}] = \frac{p[Q_\gamma^{k} \ \asymland \  Q_\beta^{j} | Q_\alpha^{i}]}{p[Q^i_\alpha |Q_\beta^{j}]},
\end{equation}
which is very similar to Bayes' theorem for conditional probability except for the presence of the operator ``$\asymland$'' instead of the conjunction ``$\land$''.

If the projectors relative to questions commute\footnote{This conditions is satisfied if the orthomodular lattice associated to the set of measurements is distributive.} additional simplifications can be applied \cite{BeltramettiCassinelli}.
In this case we have $P_\beta^{j} P_\gamma^{k} = P_\gamma^{k} P_\beta^{j}$  that, together with Eq.~\eqref{eq:3Q} gives
\begin{multline}
p[Q_\gamma^{k} | Q_\alpha^{i} \ \asymland \  Q_\beta^{j}] =\frac {tr[P_\beta^{j} \rho_\alpha P_\beta^{j} P_\gamma^{k}]}{tr[\rho_\alpha P_\beta^{j}]}  
= \frac {tr[\rho_\alpha P_\beta^{j} P_\gamma^{k}]}{tr[\rho_\alpha P_\beta^{j}]} = \\
=  \frac{p[Q_\gamma^{k} \land Q_\beta^{j} | Q_\alpha^{i}]}{p[Q_\beta^{j} | Q_\alpha^{i}]} \equiv p[Q_\gamma^{k} | Q_\alpha^{i} \land Q_\beta^{j}] ,
\end{multline}
which is nothing else the Bayes' theorem with ``$\asymland$'' equivalent to ``$\land$''.
We note that the demonstration of the Bayes' theorem provided by Cox and Jaynes \cite{Cox1946,Jaynes} is based on the assumption that the conditional probability $p[Q_\gamma^{k} \land Q_\beta^{j} | Q_\alpha^{i}]$ can be written as function of $p[Q_\gamma^{k} | Q_\alpha^{i} \land Q_\beta^{j}] $ and $p[Q_\beta^{j} | Q_\alpha^{i}]$ \cite{Cox1946,Fine,Jaynes}. 
This is clearly not true for the general case described by Eq.~\eqref{eq:3Q} where the conjunction `$\land$' between non-compatible questions is not defined \cite{Ballentine1986}.
In the other way around, we note also that Bayes' theorem can be easily derived when distributive property is assumed valid \cite{DAgostini}, which underlines once more the important role of distributivity.

\section{An example: the Young's slits} \label{sec:slits}

\subsection{Young's slits in the Relational Quantum Mechanics framework}

\begin{figure}
\centering
\includegraphics[width=0.7\textwidth]{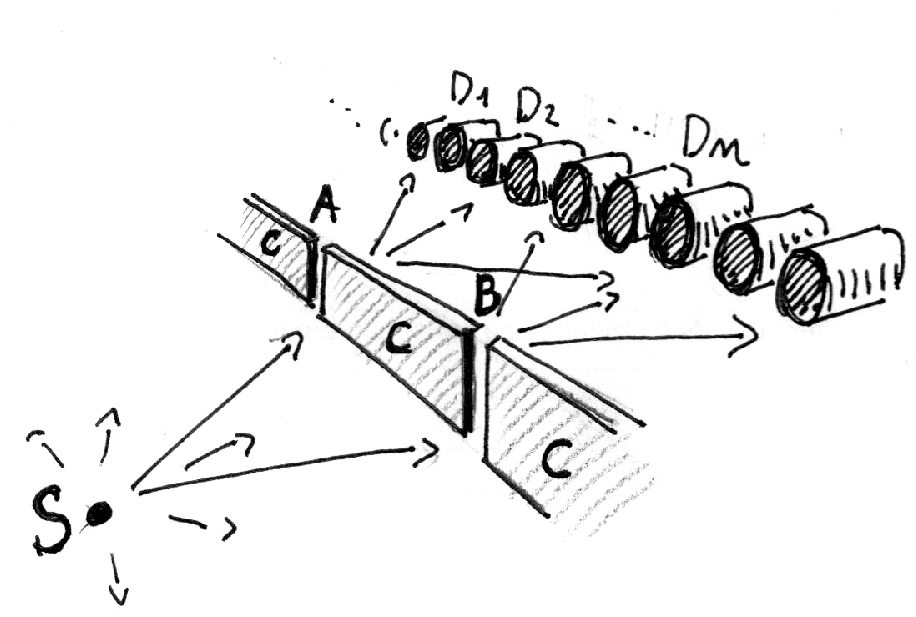}
\caption{Artistic scheme of the Young's slit experiment.}
\label{fig:slits}
\end{figure}

To see the implication of the previous formulas and their interpretation, we consider the classical example of the a `Young's slits'-like experiment, extensively considered in the literature (see e.g. \cite{Feynman1951,BeltramettiCassinelli,Hughes}).

We consider here the ideal experiment represented in Fig.~\ref{fig:slits} where a particle is emitted by a source $S$ and detected by a series of detectors $D_0, \ldots,  D_N$ after passing through a pair of slits, $A$ and $B$. 
We assume that the walls of the slits themselves, indicated by $C$, can detect the arrival of particles.

The initial state, i.e. the initial question $Q_\alpha^{i} = Q_\alpha^{S}$, corresponds to the emission or not of a particle from the source $S$.
The detection in proximity of the slits correspond to the questions $Q_\beta^{j}$ with $j=\{A,B,C\}$ for the passage through the slits $A$,$B$ or the hitting of the slits walls $C$.
The final set of questions corresponds to the detection on the the different detectors on the screen with $Q_\gamma^{k}$ with $k=\{D_0, \ldots,  D_N\}$.
Each set of questions $\{Q_\beta^{A},Q_\beta^{B},Q_\beta^{C}\}$ and $\{Q_\gamma^{D_0},\ldots,Q_\gamma^{D_N}\}$ are complete and corresponds to complete bases of Hilbert space $\{\ket{A_\beta},\ket{B_\beta},\ket{C_\beta}\}$ and $\{\ket{D_0}_\gamma, \ldots ,\ket{D_N}_\gamma \}$.

We consider two different cases. 
In the first one we consider the possibility to distinguish the passage of a particle through $A$ or $B$ (for ex. following the trajectory of the particle if the particle is a ball or another macroscopic object) before being detected on the screen. 
In the second case, we consider that only the information from the slits wall $C$ is available and we cannot access to the information of the passage through  $A$ or $B$. 
In both case we will study the detection from one particular detector only $D=D_n \in \{D_0, \ldots,  D_N\}$.

\hfill

In the first case, because we can distinguish a path between $A$ or $B$, we are considering a detection that correspond to the question structure $(Q_\gamma^{D} \ \asymland \  Q_\beta^{A}) \lor (Q_\gamma^{D} \ \asymland \  Q_\beta^{B})$: we want to determine the probability to detect a particle from $D$ passed through $A$ or to detect a particle from $D$ passed throuhg $B$.
We are only interested to the detection in $D$ (from here the operator ``$ \lor''$) but we could in principle detect from which slit the particle passed (from here the operator ``$ \asymland$'').
The probability we are interested is then $p[(Q_\gamma^{D} \ \asymland \  Q_\beta^{A}) \lor (Q_\gamma^{D} \ \asymland \  Q_\beta^{B}) | Q_\alpha^{S}]$
where $Q_\alpha^{S} = $ \textit{True} correspond to a particle emission from the source.
Detection of a passage through $A$ or $B$ correspond to mutually exclusive questions $Q_\beta^{A}$ and $Q_\beta^{B}$ where $Q_\beta^{A} \land Q_\beta^{B} = \emptyset$.
Any combination with these questions, $(Q_\gamma^{D} \ \asymland \  Q_\beta^{A})$ and $(Q_\gamma^{D} \ \asymland \  Q_\beta^{B})$ in our case, is also mutually exclusive as well.\footnote{Because we can distinguish the trajectories, each pair $(Q_\gamma^{D} \ \asymland \  Q_\beta^{A})$ and $(Q_\gamma^{D} \ \asymland \  Q_\beta^{B})$ can be considered as two distinct detectors to which correspond the questions $Q^{DA}$ and $Q^{DB}$ with $Q^{DA} \land Q^{DB} = \emptyset$.}
Applying Eq.~\eqref{eq:ax3}, we have then
\begin{multline}
p[(Q_\gamma^{D} \ \asymland \  Q_\beta^{A}) \lor (Q_\gamma^{D} \ \asymland \  Q_\beta^{B}) | Q_\alpha^{S}] = 
p[Q_\gamma^{D} \ \asymland \  Q_\beta^{A} | Q_\alpha^{S}] +  p[Q_\gamma^{D} \ \asymland \  Q_\beta^{B} | Q_\alpha^{S}] = \\
= | \braket{D_\gamma |  P_\beta^{A} |S_\alpha} |^2 + | \braket{D_\gamma |  P_\beta^{B} |S_\alpha} |^2 = \\
= | \braket{D_\gamma |A_\beta}\braket{A_\beta | S_\alpha} |^2 +| \braket{D_\gamma |B_\beta}\braket{B_\beta | S_\alpha} |^2. \label{eq:pAB}
\end{multline}
The final probability is the sum of the individual probabilities (composed each by the square product of different terms) to pass through the slit $A$ and the slit $B$.

\hfill

In the second case we consider that we cannot distinguish in principle the passage through one specific slit.  
The non-detection on the walls $C$ after an emission from $S$ corresponds to the certainty of the passage of a particle through $A$ or $B$. 
Symbolically it means that $\neg  Q_\beta^{C} \equiv Q_\beta^{C^\bot} \equiv Q_\beta^{A} \lor Q_\beta^{B}$. 
In this case, using Eq.~\eqref{eq:3Qbis}, the probability to detect a particle from $D$ is  
\begin{equation} 
p[Q_\gamma^{D} \ \asymland \  (Q_\beta^{A} \lor Q_\beta^{B}) | Q_\alpha^{S}] = | \braket{ D_\gamma | P^{A \lor B}_\beta |  S_\alpha}|^2  \end{equation} 
with $P^{A \lor B}_\beta$ the projector operator corresponding to the positive answer to the question \textit{``Is the particle passed through $A$ or $B$?''}. 

Considering that $\{ \ket{j_\beta}\}$ basis is orthonormal (formed by mutual exclusive questions), we have that $\braket{j_\beta | k_\beta}  =  0 = P^j_\beta P^k_\beta$ for $j \ne k$ and then \cite{Piron1972}
\begin{equation}
P^{A \lor B}_\beta = I - P_\beta^{A^\bot \land B^\bot} 
= I - (I -P_\beta^{A})(I-P_\beta^{B}) = P_\beta^{A} + P_\beta^{B}.
\end{equation}
Using this last equation we have that the final probability is
\begin{multline}
 p[Q_\gamma^{D} \ \asymland \   (Q_\beta^{A} \lor Q_\beta^{B}) | Q_\alpha^{S}]  = 
| \braket{D_\gamma |  (P_\beta^{A} + P_\beta^{B})|S_\alpha}|^2=  \\
 = | \braket{D_\gamma |A_\beta}\braket{A_\beta | S_\alpha} +  \braket{D_\gamma |B_\beta}\braket{B_\beta|S_\alpha}|^2.  \label{eq:pAorB}
 \end{multline}
Differently to the discernible case (Eq.~\eqref{eq:pAB}), we have here the square of the sum of two terms with the possibility of interference terms.
We have then the well known property of probability calculation of QM without introducing additional postulates or axioms.\footnote{An alternative demonstration of the derivation of the superposition property can be found in Ref.~\cite{BeltramettiCassinelli} (Section 26.3).}

In this example we considered only two slits, but it can easily extended to a general case of $N_s$ slits. 
The ensemble of slits correspond to a set of mutually exclusive questions $\{Q_\beta^A, Q_\beta^B, Q_\beta^E, Q_\beta^F, \ldots \}$ with an additional question corresponding to the ensemble of slit walls $Q_\beta^{C^\bot} \equiv Q_\beta^{A} \lor Q_\beta^{B} \lor Q_\beta^{E} \lor Q_\beta^{F} \lor \ldots$ for which Eqs.~\eqref{eq:pAB} and \eqref{eq:pAorB} can be easily generalized.

\subsection{Non-distributivity of the probability argunents}

As presented in the previous section, the interference effect on the final probability by one of the detectors $D_0, D_1, \ldots, D_N$ is directly related to the possibility to distinguish from where the particle passed, i.e. the possibility or not to access to the information relative to the passage through a particular slit.
Because of the first RQM postulate, this information can be in fact intrinsically limited.
The possibility to distinguish or not the particle path correspond to different conditional probability expressions, Eqs.~\eqref{eq:pAB} and \eqref{eq:pAorB} respectively.
Their equivalence cannot be assumed a priori in the slits experiment example because the distributive property, habitually assumed as valid as in common (classical) cases, is not justified here and can be violated, with in our specific case
\begin{equation}
Q_\gamma^{D} \ \asymland \   (Q_\beta^{A} \lor Q_\beta^{B}) \ne (Q_\gamma^{D} \ \asymland \  Q_\beta^{A}) \lor (Q_\gamma^{D} \ \asymland \  Q_\beta^{B}).
\end{equation}
Distributive property is non-valid in question/experiment lattices involving a limited amount of information and it cannot be assumed by default, as presented in a specific example in the Appendix.
If distributive property is not given for granted, we have to distinguish the different cases as in the example of Young's slits experiment.
Different cases correspond to different expressions of the probability function $p$, which is uniquely defined by Eqs.~\eqref{eq:ax}. 
What it changes is the proposition combination (question combination) for which we are calculating the probability.

Because of the contradiction with daily experience phenomena where the distributive property is valid, its abandon is, to the opinion of the author, the largest intellectual difficulty in this framework to reconstruct QM formalism from RQM postulates.
However the other advantage of RQM formulation is the possibility to transpose the series of questions/measurements to a set of propositions that can be studied then in the context of proposition logic.
In the case of the slits experiment, once a particle has been emitted and detected, the sentence \textit{``The particle passed through the slit A OR the particle passed through the slit B''} is always true.
In opposite the sentence \textit{``The particle passed through both slits A and B''}, equivalent to $Q_\beta^{A} \land Q_\beta^{B}$ is always \textit{False} because $Q_\beta^{A}$ and $Q_\beta^{B}$ are in fact compatible and mutually exclusive with $Q_\beta^{A} \land Q_\beta^{B} = \emptyset$ ($P^A_\beta P^B_\beta = 0$).
Finally, the sentences \textit{``The particle passed through the slit A''} and \textit{``The particle passed through the slit B''} are completely indeterminate if we do not measure where the particle passed through.
This is similar to an analog case discussed by Aristotle (reported in Ref.~ \cite{DallaChiara}) well before the formulation of quantum mechanics.  The sentences \textit{`Tomorrow there will be a sea-battle''} and its negation \textit{`Tomorrow there will not be a sea-battle''} have today no definite truth-value (they are indeterminate), but their disjunction \textit{``Either tomorrow there will be a sea-battle or tomorrow there will not be a sea-battle''} is true. As discussed in Refs.~\cite{DallaChiara,Auletta}, this structure can be related to the Łukasiewicz three-valued logic approach \cite{Bergmann} where in addition to the \textit{True} and \textit{False} logical values for propositions, an \textit{Indeterminate} or \textit{Unknown} third value is considered.

\section{Conclusions} \label{sec:conclusions}
In conclusion we demonstrated that from a rigorous definition of the conditional probability for the possible outcomes of different measurements,  the Born's rule and the more general rule with the density operator naturally emerges from the first two postulates of Relational Quantum Mechanics by the use of the Gleason's theorem.
Thus this eliminates the need of introducing a third postulate.
In the context of Relational Quantum Mechanics, the approach of the presented work is an alternative to the method adopted by Höhn \cite{Hohn2017,Hohn2017a,Hohn2017b}. 
There, the complex field of the reconstructed Hilbert space is univocally determined, but it requires additional postulates and the assumption of the symmetric properties of the probability function $p$. 
This is not the case in the presented work where symmetric properties of $p$ are derived together with the Born’s rule.

The presence or not of interference terms is demonstrated to be related to the precise formulation of the conditional probability where distributive property cannot be taken for granted.
In the specific case of Young's slits, the violation of distributive property corresponds to a difference between the probability \textit{to detect a particle that passed through the pair of slits A and B} ($Q_\gamma^{D} \ \asymland \   (Q_\beta^{A} \lor Q_\beta^{B})$) and the probability  \textit{to detect a particle that passed through slit A OR to detect a particle that passed through slit B} ($(Q_\gamma^{D} \ \asymland \  Q_\beta^{A}) \lor (Q_\gamma^{D} \ \asymland \  Q_\beta^{B})$).
Even if the two propositions are apparently equivalent in common language, there is a important difference linked to the possibility or not to determine the particle passage through a particular slit bringing to the disequality $Q_\gamma^{D} \ \asymland \  Q_\beta^{A}) \lor (Q_\gamma^{D} \ \asymland \  Q_\beta^{B}) \ne Q_\gamma^{D} \ \asymland \   (Q_\beta^{A} \lor Q_\beta^{B})$ in the possible probability arguments.
The already well known non-validity of distributive property in the logical structure of ``yes/no'' measurements manifests here once more in the formulation of the probability function $p$.

\begin{acknowledgements}
I would like to thank B. Delamotte, R. Grisenti, E. Lamour, M. Marangolo, V. Parigi, C. Prigent and L. Simons, that helped me, with many discussions on the interpretation of quantum mechanics and probability, to produce the present work.
In particular I would like to thank C. Rovelli for inciting and encouraging the publication of this work, M. Romanelli for the attentive readings of the different version of the article and for the many discussions, and M. Lupi who, with a simple gift, made me ask some questions that seeded this work.
\end{acknowledgements}

\section*{Appendix A: ``Yes/No'' measurements and orthomodular lattices}

\subsection*{A1 A simple case of distributive and orthocomplemented lattice}

To present orthomodular lattices, their properties and their link to a set of ``yes/no'' experiments, we will consider two simple cases: a classic case relative to the measurement of the size of eggs using two calibers (an example introduced originally by Piron \cite{Piron1972}), and a case of measurement of a spin-1 particle with a Stern-Gerlach apparatus. 

\begin{figure}
\centering
\includegraphics[width=0.5\textwidth]{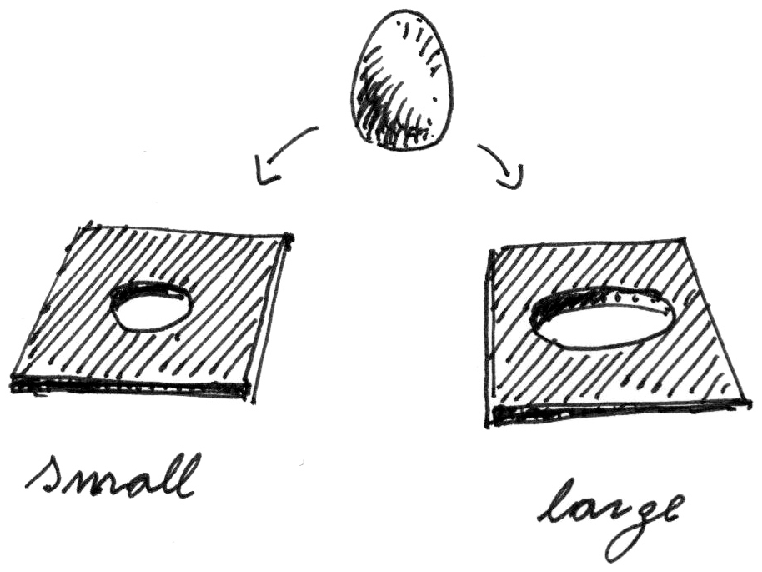} \hspace{1cm}
\includegraphics[height=0.25\textheight]{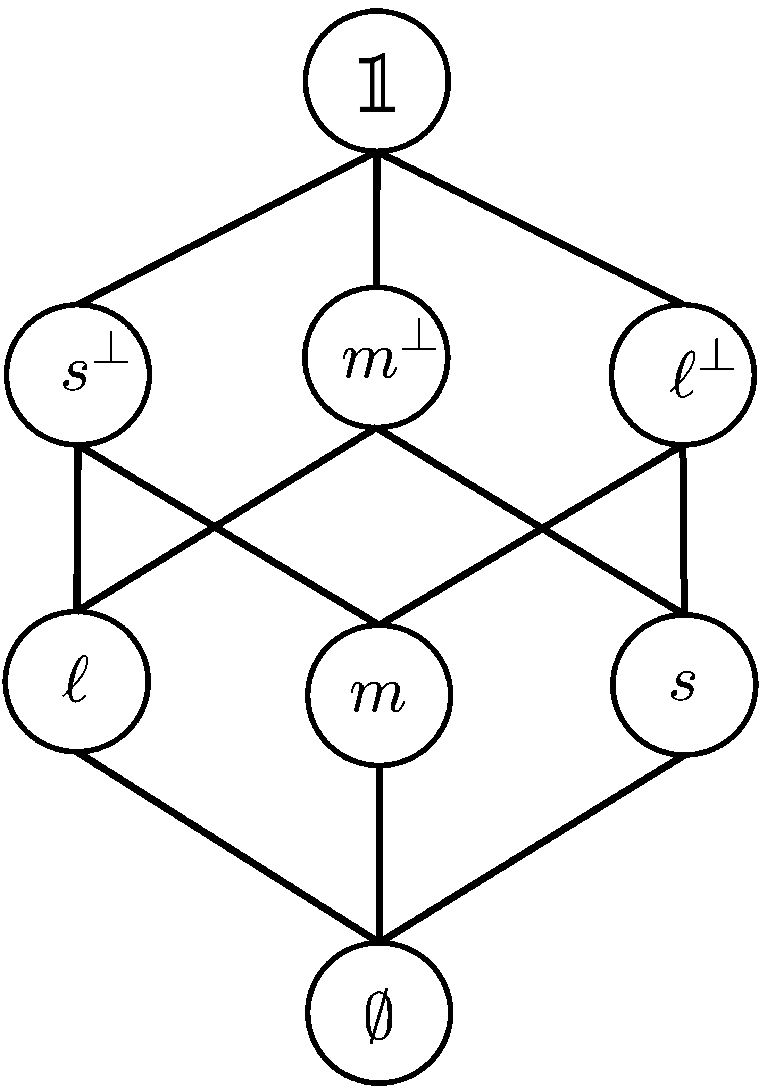}
\caption{Left: artistic scheme of a egg size measurement with two calibers.
Right: graph of the orthocomplemented lattice corresponding to the possible outcomes on the egg size.}
\label{fig:egg}
\end{figure} 

In the case of the egg measurement, the questions $\{Q_\alpha \}$ associated to these measurements are composed by $Q^s =$``\textit{is the egg small?}'', which positive answer correspond to an egg passing through the small hole with diameter $d_s$, and by $Q^\ell =$ ``\textit{is the egg large?}'' corresponding to an egg not passing through the large hole with diameter $d_\ell > d_s$ (see Fig.~\ref{fig:egg} left). 
From the combination of these two questions we can deduce three sizes: ``\textit{s}'' \textit{small} ($Q^s$), ``$\ell$'' \textit{large} ($Q^\ell$) and ``\textit{m}'' \textit{medium} from the combination $Q^m \equiv \lnot Q^s \land \lnot Q^\ell$.

The different answers (positive or negative) can be interpreted as propositions that can be true or false and are related to each other.
As example, if we have a positive answer to the question $Q^s$, this corresponds to the proposition ``\textit{the egg is small}'' $ \equiv s$.
If the we have a negative answer (positive to its negation $\lnot Q^s$), this corresponds to the proposition ``\textit{the egg is not small}'' $ \equiv s^\bot$, where ``$\bot$'' indicates complementation (orthogonality) of negative answers with respect to positive answers. 
In addition to the complementation ``$\bot$'' we can introduces the operations of conjunction ``$\land$'' (``AND''), disjunction ``$\lor$'' (``OR'') between the propositions $s,m,\ell, s^\bot,m^\bot,\ell^\bot$.

The different answers/propositions and their relation constitute a logical structure of a   \textit{lattice}.
As example, a negative answer to the question $Q^\ell$, i.e. $\lnot Q^\ell$, implies a positive answer to $Q^m$ or $Q^s$. 
In this specific example, we have $\ell^\bot \implies m \lor s$ (``not large implies medium or small'').
This means that for every egg measurement result giving as answers ``\textit{the egg is medium}''  or ``\textit{the egg is small}'', we are sure that the proposition ``\textit{the egg is not large}'' is true.
This order relation (related to the logical implication $\implies$) can be indicated by ``$\succeq$''. 
In the specific example considered, we have that $\ell^\bot \succeq m$, $\ell^\bot \succeq s$ and $\ell^\bot \succeq m \lor s$.
This order relation is represented by a graph (Fig.~\ref{fig:egg} right) by descending lines and where the elements $i \land j$ and $i \lor j$ can be determined joining the two lower or upper lines, respectively, generating from the elements $i$ and $j$.

Due to the question/answer basic structure, each element $i$ of the partially ordered set so defined has a complement $i^\bot$ with the properties $i \lor i^\bot = \mathbb{1}$ and $i \land i^\bot = \emptyset$.
$\mathbb{1},  \emptyset$ are also part of the set with $\mathbb{1}$ corresponding to all type of eggs and $\emptyset$ to no one.
Because of these properties, our partially ordered set is actually a orthocompletemented lattice\footnote{The considered poset is a lattice because for any elements $i,j$, we can define the element combinations $i \land j$ and $i \lor j$ are part also of the considered ensemble of propositions (answers).
It is a orthocomplemented lattice because for any element $i$, there is an element $i^\bot$ with the properties $i \lor i^\bot = \mathbb{1}$ and $i \land i^\bot = \emptyset$ and $(i^\bot)^\bot$=i and $i \succeq j \implies j^\bot \succeq i^\bot$.} $\mathcal{L}$.
Another important property of the algebra of the lattice defined by this series of measurement is the distributivity (Eq.~\eqref{eq:distributivity}.

Another simple example is the case of a spin-1 particle passing through a Stern-Gerlach apparatus discussed in Sec.~\ref{sec:1RQM}.
The complete set of the answers to the questions $\{Q_V^-\}$ has the same structure as the eggs measurement.

\subsection*{A2 A more complex case; orthomodularity and distributivity}

\begin{figure}
\centering
\includegraphics[height=0.31\textheight]{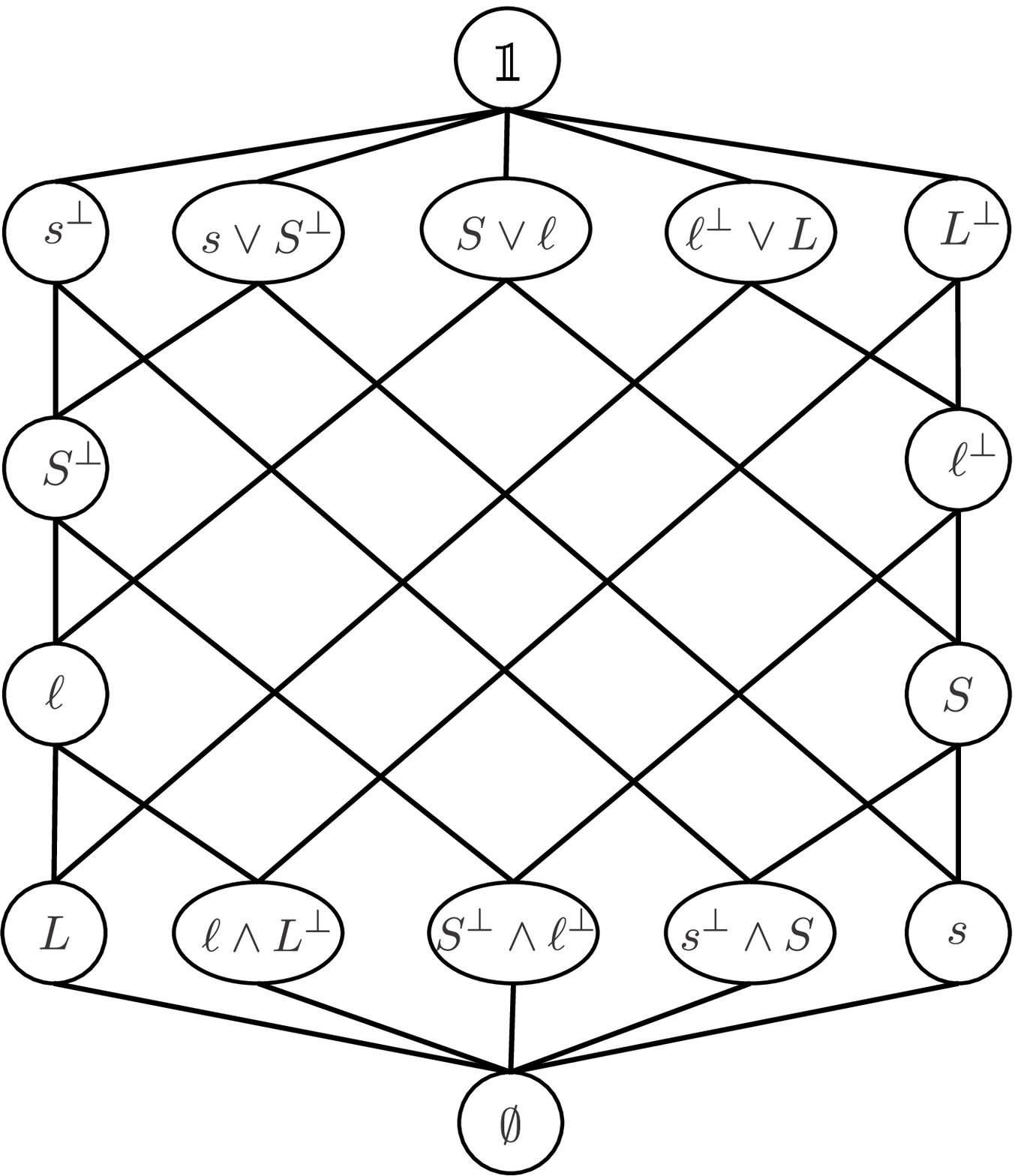} \hspace{1cm}
\includegraphics[height=0.25\textheight]{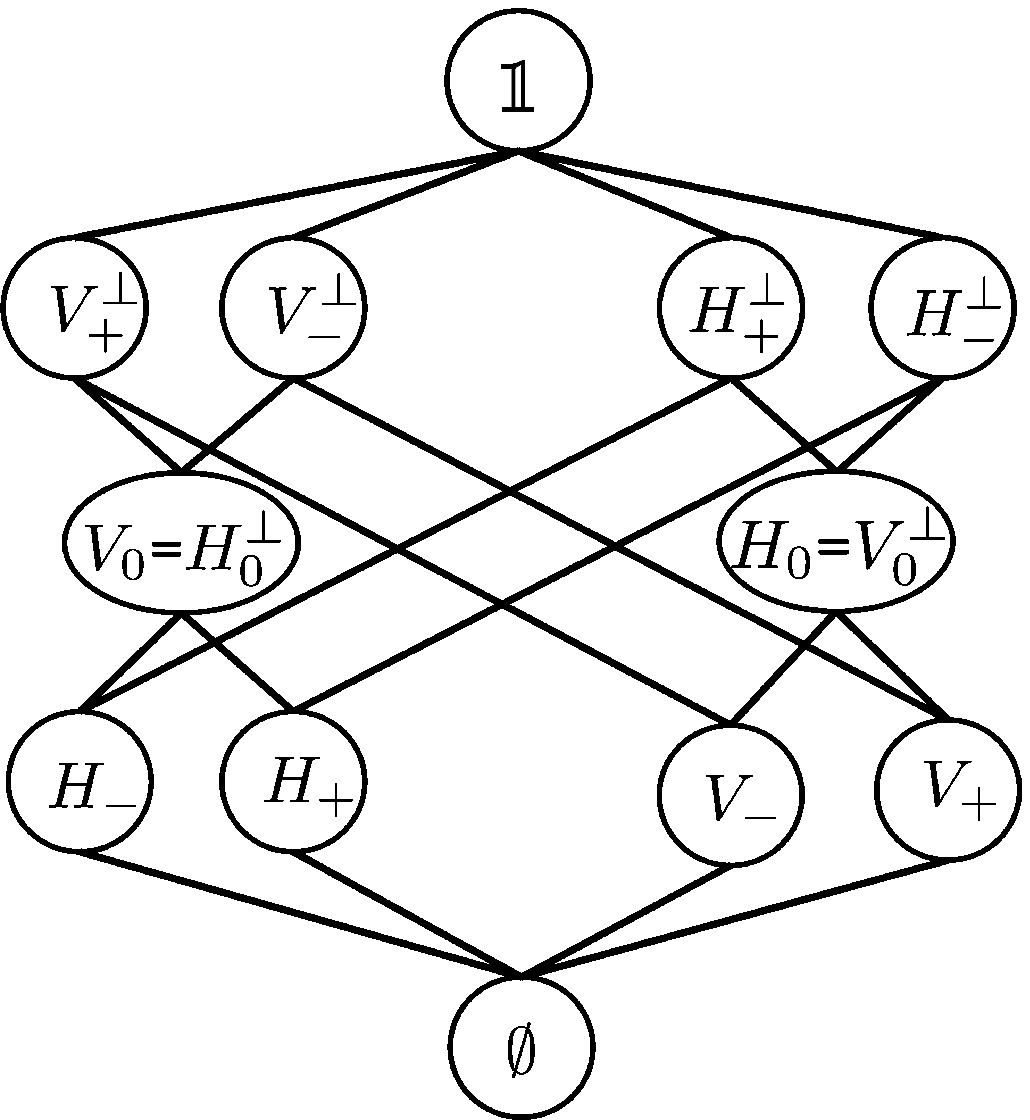}
\caption{Graphs of the orthocomplemented lattice corresponding to the measurement of egg size with four different calibers (left) and to the measurement with two Stern-Gerlach apparatus, one vertical and one horizontal.}
\label{fig:2sets}
\end{figure} 

In the classic case of the egg-size determination, a new set of questions $\{ Q_\beta \}$ corresponds to a new set of calibers with different diameters.
We can consider for example, slightly larger holes $d_S,d_L$ with $d_s<d_S<d_\ell<d_L$ that correspond to $Q^S$ and $Q^L$ questions and that generate a new set of possible five measurable sizes.\footnote{As in the previous example, the measurement ``\textit{m}'' medium come from the combination $Q^m \equiv \lnot Q^s \land \lnot Q^\ell$), with the set of questions $Q^s,Q^S, Q^\ell,Q^L$ different sizes can be defined as e.g. \textit{``not very small''} corresponding to $\lnot Q^s \land Q^S$ the intermediate size between the two smaller diameters (the egg pass through the $S$ hole but not the $s$ one.}
The corresponding lattice is represented in Fig.~\ref{fig:2sets} (left). 
It is more complex but still distributive.\footnote{Similar example of such a lattice can be found in Ref.~\cite{Hughes}.}
As discussed by Piron \cite{Piron1972,Piron} and Beltrametti and Cassinelli \cite{BeltramettiCassinelli}, this reflects the fact that the two set of questions (two pairs of calibers) are compatible with each other and that the final measurement output does not depend on the order on which the two pairs of calibers are implemented. 

In the case of the spin-1 particle, in addition to a vertically Stern-Gerlach apparatus, we can consider a horizontally oriented one. 
This new measurement apparatus is related to a new set of questions $\{ Q_H \}= Q_H^+, Q_H^0, Q_H^-$ that corresponds to new possible measurement results $\{ H_+, H_0, H_-, H^\bot_+, H^\bot_0, H^\bot_-\}$.
When the two sets of measurements are considered together, they form a non-trivial but relatively simple lattice represented in Fig.~\ref{fig:2sets} (right).
The perpendicularity of the two apparatus axes determines the fact that a particle giving the measurement result $H_0$ in the horizontal apparatus ($V_0$) corresponds to a measurement $V_+^\bot \lor V_-^\bot$  in the vertical apparatus ($H_+^\bot \lor H_-^\bot$ in the horizontal one).
The most important characteristic is that the lattice is no more distributive.
For example, if we consider the possible outcomes $H_+,V_-, V_-^\bot$, we have that
\begin{equation}
H_+ \lor (V_- \land V_-^\bot) \ne (H_+ \lor V_-) \land (H_+ \lor V_-^\bot).
\end{equation}
In fact\footnote{In the right term, $V_- \land V_-^\bot = \emptyset$ because of the complementarity and then $H_+ \lor \emptyset = H_+$.
In the left term $H_+ \lor V_- = \mathbb{1}$ (the is no other common elements) and $H_+ \lor V_-^\bot =  V_-^\bot$ because of the lattice hierarchy, so the right term is equal to $\mathbb{1} \land V_-^\bot = V_-^\bot$.}  the left term is equal to $H_+$ and the right term is equal to $V_-^\bot$.

Additional examples of non-distributive orthomodular lattices can be found in the literature, as the lattice corresponding to the measurement of polarised light \cite{Piron1972} or a quantised vector in the real space \cite{Hughes}.

As discussed by \cite{Piron1972}, the distributive property is related to the logical structure of the orthomodular lattice $\mathcal{L}$ corresponding to the ensemble of the measurements.
For two elements (``yes/no'' measurements) $i,j \in \mathcal{L}$, we have that 
\begin{equation}
``i \; \textit{True}'' \; \text{or} \; ``j \; \textit{True}'' \implies ``i \lor j\; \textit{True}'',
\end{equation} 
but 
\begin{equation}
 ``i \lor j\; \textit{True}'' \implies``i \; \textit{True}'' \; \text{or} \; ``j \; \textit{True}'', \label{eq:notimp}
\end{equation} 
is valid only if the lattice $\mathcal{L}$ is also distributive, which is not the case for measurements reaching the limit of the extractable information, i.e, the quantization aspects \cite{Birkhoff1936,VonNeumann,Piron1972} (a simple presentation of this aspect in a Hilbert space can be found in \cite[p.~37]{DallaChiara} and in  \cite[p.~190]{Hughes}).


\bibliographystyle{ams_modified}


\begin{thebibliography}{10}

\bibitem{Bohr1913}
N. Bohr,
\newblock {\itshape On the constitution of atoms and molecules},
\newblock Philos. Mag.
\newblock \textbf{ 26} (1913), 476--502.

\bibitem{Feynman1967}
R. Feynman,
\newblock {\itshape The Character of Physical Law},
\newblock M.I.T. Press,
\newblock 1967.

\bibitem{Auletta}
G. Auletta,
\newblock {\itshape Foundations and Interpretation of Quantum Mechanics: In the
  Light of a Critical-historical Analysis of the Problems and of a Synthesis of
  the Results},
\newblock World Scientific,
\newblock 2001.

\bibitem{Hardy2001}
L. Hardy,
\newblock {\itshape Quantum theory from five reasonable axioms},
\newblock
\newblock arXiv preprint quant-ph/0101012.

\bibitem{Clifton2003}
R. Clifton, J. Bub, and H. Halvorson,
\newblock {\itshape Characterizing quantum theory in terms of
  information-theoretic constraints},
\newblock Found. Phys.
\newblock \textbf{ 33} (2003), 1561--1591.

\bibitem{Grinbaum2007}
A. Grinbaum,
\newblock {\itshape Reconstruction of quantum theory},
\newblock Br. J. Philos. Sci.
\newblock \textbf{ 58} (2007), 387--408.

\bibitem{Grinbaum2007a}
A. Grinbaum,
\newblock {\itshape Reconstructing instead of interpreting quantum theory},
\newblock Philos. Sci.
\newblock \textbf{ 74} (2007), 761--774.

\bibitem{Dakic2009}
B. Dakic and C. Brukner,
\newblock {\itshape Quantum theory and beyond: Is entanglement special?},
\newblock
\newblock arXiv preprint arXiv:0911.0695.

\bibitem{Hohenberg2010}
P.~C. Hohenberg,
\newblock {\itshape \textit{Colloquium} : An introduction to consistent quantum
  theory},
\newblock Rev. Mod. Phys.
\newblock \textbf{ 82} (2010), 2835--2844.

\bibitem{Chiribella2011}
G. Chiribella, G.~M. D’Ariano, and P. Perinotti,
\newblock {\itshape Informational derivation of quantum theory},
\newblock Phys. Rev. A
\newblock \textbf{ 84} (2011), 012311.

\bibitem{Fuchs2013}
C.~A. Fuchs and R. Schack,
\newblock {\itshape Quantum-bayesian coherence},
\newblock Rev. Mod. Phys.
\newblock \textbf{ 85} (2013), 1693--1715.

\bibitem{Masanes2013}
L. Masanes, M.~P. Müller, R. Augusiak, and D. Pérez-García,
\newblock {\itshape Existence of an information unit as a postulate of quantum
  theory},
\newblock Proc. Natl. Acad. Sci. U.S.A.
\newblock \textbf{ 110} (2013), 16373--16377.

\bibitem{Barnum2014}
H. Barnum, M.~P. Müller, and C. Ududec,
\newblock {\itshape Higher-order interference and single-system postulates
  characterizing quantum theory},
\newblock New J. Phys.
\newblock \textbf{ 16} (2014), 123029.

\bibitem{Dickson2015}
M. Dickson,
\newblock {\itshape Reconstruction and reinvention in quantum theory},
\newblock Found. Phys.
\newblock \textbf{ 45} (2015), 1330--1340.

\bibitem{Auffeves2016}
A. Auffèves and P. Grangier,
\newblock {\itshape Contexts, systems and modalities: A new ontology for
  quantum mechanics},
\newblock Found. Phys.
\newblock \textbf{ 46} (2016), 121--137.

\bibitem{Friedberg2018}
R. Friedberg and P.~C. Hohenberg,
\newblock {\itshape What is quantum mechanics? a minimal formulation},
\newblock
\newblock Found. Phys.

\bibitem{Rovelli1996}
C. Rovelli,
\newblock {\itshape Relational quantum mechanics},
\newblock Int. J. Theor. Phys.
\newblock \textbf{ 35} (1996), 1637--1678.

\bibitem{Brukner1999}
C. Brukner and A. Zeilinger,
\newblock {\itshape Operationally invariant information in quantum
  measurements},
\newblock Phys. Rev. Lett.
\newblock \textbf{ 83} (1999), 3354--3357.

\bibitem{Brukner2002}
C. Brukner and A. Zeilinger,
\newblock {\itshape Young's experiment and the finiteness of information},
\newblock Phil. Trans. R. Soc. A
\newblock \textbf{ 360} (2002), 1061--1069.

\bibitem{Fuchs2002}
C.~A. Fuchs,
\newblock {\itshape Quantum mechanics as quantum information (and only a little
  more)},
\newblock
\newblock arXiv preprint quant-ph/0205039.

\bibitem{Barrett2007}
J. Barrett,
\newblock {\itshape Information processing in generalized probabilistic
  theories},
\newblock Phys. Rev. A
\newblock \textbf{ 75} (2007), 032304.

\bibitem{Brukner2009}
C. Brukner and A. Zeilinger,
\newblock {\itshape Information invariance and quantum probabilities},
\newblock Found. Phys.
\newblock \textbf{ 39} (2009), 677--689.

\bibitem{Hohn2017}
P.~A. Höhn and C.~S.~P. Wever,
\newblock {\itshape Quantum theory from questions},
\newblock Phys. Rev. A
\newblock \textbf{ 95} (2017), 012102.

\bibitem{Hohn2017a}
P. Höhn,
\newblock {\itshape Toolbox for reconstructing quantum theory from rules on
  information acquisition},
\newblock Quantum
\newblock \textbf{ 1} (2017), 38.

\bibitem{Hohn2017b}
P. Höhn,
\newblock {\itshape Quantum theory from rules on information acquisition},
\newblock Entropy
\newblock \textbf{ 19} (2017), 98.

\bibitem{Yang2017}
J. Yang,
\newblock {\itshape Quantum mechanics from relational properties, part {I}:
  Formulation},
\newblock
\newblock arXiv preprint arXiv:1706.01317.

\bibitem{Yang2018}
J. Yang,
\newblock {\itshape Quantum mechanics from relational properties, part {II}:
  Measurement},
\newblock
\newblock arXiv preprint arXiv:1803.04843.

\bibitem{Yang2018a}
J. Yang,
\newblock {\itshape Quantum mechanics from relational properties, part {III}:
  Path integral implementation},
\newblock
\newblock arXiv preprint arXiv:1807.01583.

\bibitem{Smerlak2007}
M. Smerlak and C. Rovelli,
\newblock {\itshape Relational epr},
\newblock Found. Phys.
\newblock \textbf{ 37} (2007), 427--445.

\bibitem{Rovelli2018}
C. Rovelli,
\newblock {\itshape ‘space is blue and birds fly through it’},
\newblock Phil. Trans. R. Soc. A
\newblock \textbf{ 376}.

\bibitem{Poulin2006}
D. Poulin,
\newblock {\itshape Toy model for a relational formulation of quantum theory},
\newblock Int. J. Theor. Phys.
\newblock \textbf{ 45} (2006), 1189--1215.

\bibitem{Brown2009}
M.~J. Brown,
\newblock {\itshape Relational quantum mechanics and the determinacy problem},
\newblock Br. J. Philos. Sci.
\newblock \textbf{ 60} (2009), 679--695.

\bibitem{vanFraassen2009}
B.~C. van Fraassen,
\newblock {\itshape Rovelli’s world},
\newblock Found. Phys.
\newblock \textbf{ 40} (2009), 390--417.

\bibitem{Grinbaum2005}
A. Grinbaum,
\newblock {\itshape Information-theoretic princple entails orthomodularity of a
  lattice},
\newblock Found. Phys. Lett.
\newblock \textbf{ 18} (2005), 563--572.

\bibitem{Birkhoff1936}
G. Birkhoff and J. {Von Neumann},
\newblock {\itshape The logic of quantum mechanics},
\newblock Ann. Math.
\newblock \textbf{ 37} (1936), 823--843.

\bibitem{VonNeumann}
J. {Von Neumann},
\newblock {\itshape Mathematical Foundations of Quantum Mechanics},
\newblock Princeton University Press,
\newblock 1955.

\bibitem{BeltramettiCassinelli}
E. Beltrametti and G. Cassinelli,
\newblock {\itshape The Logic of Quantum Mechanics},
\newblock Cambridge University Press,
\newblock 1984.

\bibitem{Hughes}
R. Hughes,
\newblock {\itshape The Structure and Interpretation of Quantum Mechanics},
\newblock Harvard University Press,
\newblock 1989.

\bibitem{DallaChiara}
M. {Dalla~Chiara}, R. Giuntini, and R. Greechie,
\newblock {\itshape Reasoning in Quantum Theory: Sharp and Unsharp Quantum
  Logics},
\newblock Springer Netherlands,
\newblock 2004.

\bibitem{Cohen}
C. Cohen-Tannoudji, B. Diu, and F. Laloë,
\newblock {\itshape Mécanique Quantique},
\newblock second edition,
\newblock Hermann,
\newblock Paris, 1996.

\bibitem{SEPprob}
A. Hájek,
\newblock {\itshape Interpretations of probability},
\newblock in The Stanford Encyclopedia of Philosophy (Winter 2012 Edition),
  ({Edward N. Zalta}, ed.),
\newblock 2012.

\bibitem{Fine}
T. Fine,
\newblock {\itshape Theories of probability: an examination of foundations},
\newblock Academic Press,
\newblock 1973.

\bibitem{Kolmogorov}
A. Kolmogorov,
\newblock {\itshape Foundations of the theory of probability},
\newblock Chelsea Pub. Co.,
\newblock 1956.

\bibitem{Jaynes}
E. Jaynes and G. Bretthorst,
\newblock {\itshape Probability Theory: The Logic of Science},
\newblock Cambridge University Press,
\newblock 2003.

\bibitem{Gudder}
S. Gudder,
\newblock {\itshape Quantum Probability},
\newblock Academic Press,
\newblock 1988.

\bibitem{Varadarajan}
V. Varadarajan,
\newblock {\itshape Geometry of Quantum Theory},
\newblock Springer New York,
\newblock 2007.

\bibitem{Cox}
R. Cox,
\newblock {\itshape Algebra of Probable Inference},
\newblock Johns Hopkins University Press,
\newblock 1961.

\bibitem{Ballentine1986}
L.~E. Ballentine,
\newblock {\itshape Probability theory in quantum mechanics},
\newblock Am. J. Phys.
\newblock \textbf{ 54} (1986), 883--889.

\bibitem{Ballentine2016}
L.~E. Ballentine,
\newblock {\itshape Propensity, probability, and quantum theory},
\newblock Found. Phys.
\newblock \textbf{ 46} (2016), 973--1005.

\bibitem{Auffeves2017}
A. Auffèves and P. Grangier,
\newblock {\itshape Recovering the quantum formalism from physically realist
  axioms},
\newblock Sci. Rep.
\newblock \textbf{ 7} (2017), 43365.

\bibitem{Gleason1957}
A.~M. Gleason,
\newblock {\itshape Measures on the closed subspaces of a hilbert space},
\newblock J. Math. Mech.
\newblock \textbf{ 6} (1957), 885--893.

\bibitem{Piron1972}
C. Piron,
\newblock {\itshape Survey of general quantum physics},
\newblock Found. Phys.
\newblock \textbf{ 2} (1972), 287--314.

\bibitem{Piron}
C. Piron,
\newblock {\itshape Foundations of quantum physics},
\newblock Benjamin-Cummings Publishing Company,
\newblock 1976.

\bibitem{Cox1946}
R.~T. Cox,
\newblock {\itshape Probability, frequency and reasonable expectation},
\newblock Am. J. Phys.
\newblock \textbf{ 14} (1946), 1--13.

\bibitem{DAgostini}
G. D'Agostini,
\newblock {\itshape Bayesian Reasoning in Data Analysis: A Critical
  Introduction},
\newblock World Scientific,
\newblock 2003.

\bibitem{Feynman1951}
R.~P. Feynman,
\newblock {\itshape The concept of probability in quantum mechanics},
\newblock Proceedings of the Second Berkeley Symposium on Mathematical
  Statistics and Probability,
\newblock University of California Press,
\newblock Berkeley, 1951.

\bibitem{Bergmann}
M. Bergmann,
\newblock {\itshape An introduction to many-valued and fuzzy logic: semantics,
  algebras, and derivation systems},
\newblock Cambridge University Press,
\newblock 2008.

\end{thebibliography}

\end{document}